\documentclass[a4paper]{article}
\usepackage[left=2.5cm, right=2.5cm, top=2.5cm, bottom=2.5cm]{geometry}

\usepackage{url}
\usepackage{hyperref}

\newcommand{\changefontsize}{\fontsize{12}{16}\selectfont}

\newcommand{\defaultfigsize}{\normalsize}

\changefontsize

\usepackage{cmbright}
\usepackage[T1]{fontenc}

\usepackage[utf8]{inputenc}
\usepackage{amssymb}
\usepackage{amsfonts}
\usepackage{amsmath}
\usepackage{fontsmpl}
\usepackage{graphicx}
\usepackage{epsfig}
\usepackage[utf8]{inputenc}
\usepackage[english]{babel}
 
\usepackage{amsthm}
\usepackage{subcaption}
\usepackage{epsfig}

\usepackage[table,xcdraw]{xcolor}  %need this for coloring rows in tables
 \usepackage{booktabs} % need for the tables
\usepackage{caption} 
\usepackage{subcaption}
\usepackage{rotating}

\usepackage[subfigure]{tocloft}

\usepackage{tikz}
\usepackage{subcaption}
\usepackage{xcolor}

\usepackage{booktabs}
\usepackage{multirow}
\usepackage{rotating}
\usepackage{forest}
\usepackage{xcolor}

\usepackage{graphicx}
\usepackage{booktabs}
\usepackage{float}

\usepackage{amsmath}

\title{My Boss is a Narcissist Bully: A Game Theoretic Approach to Stop Bullies}
\author{
    Pascal Stiefenhofer\footnote{pascal.stiefenhofer@newcastle.ac.uk}, \\
        \small{Newcastle University, Department of Economics}\\
    Cafer Deniz\footnote{c.deniz2@newcastle.ac.uk}, \\
    \small{Newcastle University Business School, and The University of Usak, Department of Economics}\\
    Liangxun Xie\footnote{l.xie8@newcastle.ac.uk}, \\
    \small{Newcastle University, Department of Economics}\\
    and Jing Qian\footnote{j.qian13@newcastle.ac.uk} \\
    \small{Newcastle University, Department of Economics}\\
}

\begin{document}	\changefontsize
	\maketitle

	\begin{abstract}	\changefontsize
	This paper investigates effective strategies for dealing with workplace bullying perpetrated by a narcissistic boss. Adopting a game-theoretic framework, we propose a three-stage sequential game with a simultaneous form game, incorporating a war of attrition in the final stage. Our findings demonstrate that victims of bullying should consistently choose to signal to escalate the situation and report bulling rather than ignore and tolerate the abusive behavior. Additionally, we explore how leveraging the narcissist's inherent fears can empower the victim in selecting the most advantageous equilibrium solution. By employing this comprehensive approach, individuals facing a narcissistic boss bullying can effectively address and mitigate these challenging circumstances.
	\end{abstract}
	

\section{Introduction}

Narcissism, marked by an excessive preoccupation with oneself and one’s personal needs, often presents through traits such as an inflated sense of self-importance, an intense focus on achievements and authority, and unwavering confidence in one's abilities \cite{Masterson1988}. These traits are frequently valued in corporate settings, potentially explaining the prevalence of narcissistic individuals in leadership roles \cite{Camm2014}. Understanding the drivers of narcissistic behavior and developing effective strategies for collaboration is essential in professional contexts, especially given that narcissistic tendencies are often linked to workplace bullying \cite{Lutgen-Sandvik2010, Grijalva2015, Hogan1988}.

Research shows that narcissistic individuals commonly engage in indirect bullying tactics—behaviors such as gossiping, withholding essential information, and undermining colleagues—as a means of maintaining control and asserting dominance \cite{Petrecca2010, Donohue2013, Yamada2010}. A particularly damaging pattern, known as "kiss up, kick down," describes the way some narcissistic leaders ingratiate themselves with superiors while mistreating subordinates \cite{Nevicka2018, Spain2019}. In hierarchical workplace settings, this dynamic can reinforce the bully’s authority while isolating or disempowering their targets, creating a hostile environment that diminishes team cohesion \cite{Rosen2016}.

Workplace bullying, broadly defined as repetitive, hostile actions driven by intent and a power imbalance, has been extensively researched across diverse organizational environments \cite{OConnell2007}. Such behavior, often motivated by a pursuit of dominance and status, yields negative outcomes for both victims and organizations, including heightened absenteeism, reduced job satisfaction, and elevated turnover, which collectively erode productivity and morale \cite{Pellegrini2002, Peets2023, Sijtsema2009}. Additionally, studies link workplace bullying to serious physical and mental health consequences, including increased stress, anxiety, and burnout \cite{Kivimaki2003, Houshmand2012, Nielsen2015, Einarsen2009}. These findings underscore the pressing need for effective strategies to address bullying, particularly when it originates from narcissistic leaders whose behavior can deeply damage workplace trust and cooperation \cite{Spain2019}.

In academic settings, bullying takes on unique characteristics due to the hierarchical and competitive nature of universities. Academic bullying often involves tactics such as undermining professional credibility, withholding resources, or isolating individuals from collaborative networks \cite{Keashly2010, Clark2013}. Senior faculty members may withhold vital information, research funding, or teaching opportunities from junior colleagues, compounding their stress and hindering career advancement \cite{Clark2013, Hutchinson2015}. Another common tactic is task-based bullying, where junior faculty or staff are overloaded with administrative duties that detract from research and teaching responsibilities, diminishing their professional standing \cite{Hutchinson2015}. Isolation and exclusion from departmental meetings or projects further compound these effects, leaving victims professionally marginalized and unsupported \cite{Lewis2004}. The “kiss up, kick down” dynamic is also prevalent in academic hierarchies, making it challenging to recognize and address this behavior, as academic bullies may appear supportive to superiors while subtly undermining those they perceive as subordinates \cite{Rayner2002}. Persistent, unfair criticism, particularly targeting untenured faculty or graduate students, can serve as a means of intimidation and control \cite{Bjorkqvist2001}. Together, these forms of bullying can severely impact mental health, productivity, and morale, highlighting the need for well-defined policies and interventions in academic institutions to curb such toxic behaviors \cite{McKay2008}.

Workplace bullying is alarmingly pervasive in academia, with widespread and lasting impacts on faculty well-being and performance. A 2019 survey by the American Association of University Professors (AAUP) revealed that 18\% of U.S. faculty experienced bullying or harassment, and 8\% reported significant detriments to their job performance and mental health \cite{AAUP2020}. In the UK, a 2018 University and College Union (UCU) report found that 40\% of academic staff faced bullying, with women and ethnic minorities disproportionately affected, highlighting the role of institutional power imbalances in exacerbating this issue. While, a recent 2023 survey by the Culture, Employment and Development of Academic Researchers Survey (CEDARS) showed that just over 20\% of academic staff experienced bullying or harassment in the last two years in the UK \cite{Cedars2023}. Likewise, a 2020 study by the Australian Council of Trade Unions (ACTU) showed that bullying is a major contributor to burnout and mental health struggles, especially for early-career researchers and adjunct staff. Collectively, these findings stress the urgent need for systemic reforms to counter workplace bullying in academia, creating healthier, more supportive environments where all faculty and staff can thrive.\footnote{University and College Union (2018). Australian Council of Trade Unions (2020). Workplace Bullying in Higher Education: Report on Academic Staff Experiences. American Association of University Professors (2019). The Annual Report on the Economic Status of the Profession.}

Despite the prevalence of bullying by narcissistic individuals in authority, research on effective victim response strategies remains limited. Structural power imbalances often make it difficult for victims to confront or report bullying, adding complexity to efforts to address these behaviors \cite{Einarsen2009}. Studies suggest that victims frequently hesitate to respond due to fear of retaliation, complicating their ability to counter bullying effectively \cite{Baillien2011, Namie2011}.

This paper aims to address this gap by examining response strategies for victims of workplace bullying, focusing specifically on bullying by narcissistic leaders. Drawing on existing literature and using a game-theoretic approach, we develop a framework to analyze victim responses within the context of narcissistic boss bullying. By modeling these interactions as a sequential game with elements of attrition, we explore how victims can leverage both passive and active responses, considering factors such as organizational support, the bully’s power, and control parameters.

Control variables that limit narcissistic behavior in the workplace focus on creating environments that address narcissists’ vulnerabilities and aversions, particularly their sensitivities around control, image, and social standing. Narcissists often react negatively to public criticism or any threat to their self-image, which makes performance reviews and constructive feedback valuable tools for curbing their dominance \cite{Kernis2005}. They also dislike situations that limit their autonomy, so implementing structured, transparent processes for team projects and decision-making can constrain their ability to manipulate others \cite{Campbell2009}. Additionally, promoting equality through rotating leadership roles or collaborative initiatives can minimize hierarchical power dynamics that narcissists often exploit for personal gain \cite{Grijalva2014}. Narcissists are especially wary of accountability measures—like peer reviews, anonymous feedback, and 360-degree assessments—that introduce transparency and reduce opportunities for unchecked self-serving behavior \cite{Brunell2017}. Furthermore, policies that standardize access to resources and recognize team achievements over individual praise can challenge their need for special treatment and unique privileges \cite{Emmons1984}. Direct confrontation about harmful behaviors, with specific examples and constructive solutions, can also encourage change by making narcissists more aware of the impacts of their actions \cite{Morf2001}. Finally, fostering inclusive, cohesive team environments and using social dynamics to reward positive collaboration can discourage narcissistic tendencies, as narcissists are sensitive to rejection and social isolation \cite{Leary2000}. Collectively, these strategies support a culture that limits the appeal of bullying and manipulative behaviors, promoting a more collaborative and respectful workplace.

The structure of the paper is as follows: Section two introduces our game-theoretic model of bullying by a narcissistic boss, setting the foundation for the subsequent analysis. Sections three and four then provide an in-depth exploration of victim strategies within environments characterized by varying levels of control. The paper concludes with a summary of key findings and implications for future research, offering valuable insights for developing organizational policies to mitigate narcissistic bullying behaviors.

\section{The Model}

In this model, there are two players denoted as $i=v,b.$ Player $v$ assumes the role of an employee, referred to as the victim, and is characterized by a concave utility function denoted as $u_{v}( ; )$. On the other hand, player $b$ embodies a narcissistic bully acting in a role of power, e.g., as the line manager to the victim. The bully is initially characterized by a convex utility function denoted as $u_{b}( ; )$, representing narcissistic preferences for successfully bullying the victim. We define a variable $x$ to represent the act of bullying, which ranges from $-1\leq x\leq 1$. Successful bullying is denoted as $s$ when $x\geq 0$, and $x$ lies within the interval $[-1,1]$. We associate a utility function $u_{i}(x;s)$ with successful bullying, representing the player's payoff in such a scenario. Conversely, unsuccessful bullying is denoted as $f$ when $x\leq 0$, and $x$ lies within the interval $[-1,1]$. In this case, the payoff function is denoted as $u_{i}(x;f)$. It is noteworthy that when $x=0$, the utility payoffs $u_{i}(x;s)$ and $u_{i}(x;f)$ intersect, meaning that the payoff for both players is the same when bullying is neither successful nor unsuccessful. We refer to this case the trivial case.

We examine a three-stage sequential conflict scenario. In the initial stage of this game-theoretic model, the victim is subjected to bullying by a narcissistic boss and must choose between two strategic responses. The first option is an ignoring strategy, denoted by \(I\), in which the victim tolerates the bullying behavior, incurring a negative emotional utility \( u_{v}(x;s) < 0 \), while the bully gains positive utility from their successful bullying, \( u_{b}(x;s) > 0 \). If no bullying occurs at this stage (\( x = 0 \)), the conflict remains trivial. Alternatively, the victim may choose to escalate, denoted by \( E \), advancing the situation to the next conflict stage. Escalation entails confronting the bully and actively addressing their misconduct.

At stage two, the bully, now confronted with the victim's escalation (E), faces two possible courses of action. Firstly, the bully may choose to withdraw (W) from the conflict, acknowledging the unsuccessful bullying attempt and experiencing a negative utility outcome $u_{b}(x;f)<0$ while the victim sense of victory yields a positive payoff from the narcissist’s unsuccessful bullying attempt, $u_{v}(x;f)>0$. Alternatively, the bully can opt to escalate the situation further and choose strategy $E$, deferring the resolution to a third-party judgment panel. In this case, the conflict transforms into a game of attrition, wherein both the bully and victim simultaneously choose their actions without knowledge of each other's decisions.

\begin{figure}[h!]
	\centering
	\includegraphics[scale = 0.5]{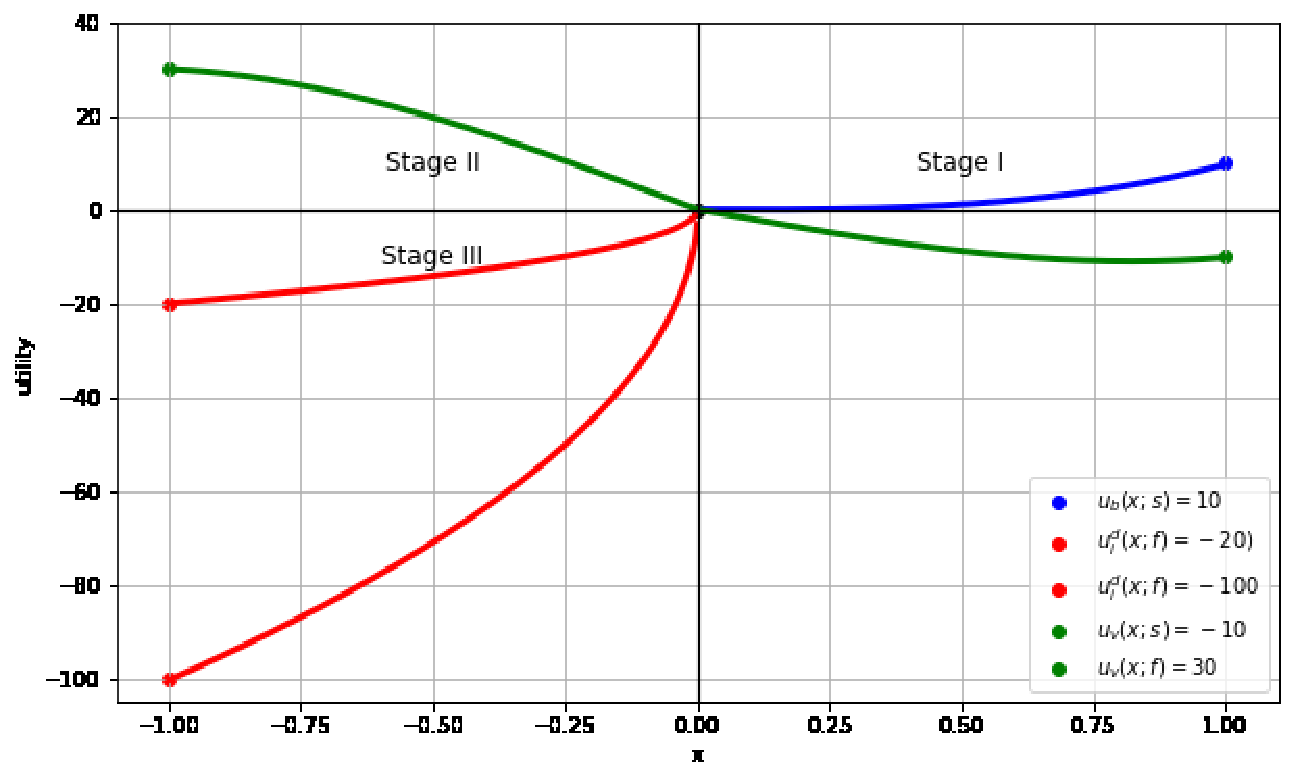}
	\caption{Main Model without control parameter: Utility payoffs of the three-stage sequential game: Stage I $u_{v}(x;s)=-10$ and $u_{b}(x;s)=10$. Stage II $u_{v}(x;f)=30$ and $u_{b}(x;f)=-30$. Stage III  $u_{i}(x;f)=-20$ if  stategy $(R,r)$ and $u_{i}(x;f)=-100$ if any mutual destruction based strategy is chosen. }
	\label{fig: main stable ut}
\end{figure}

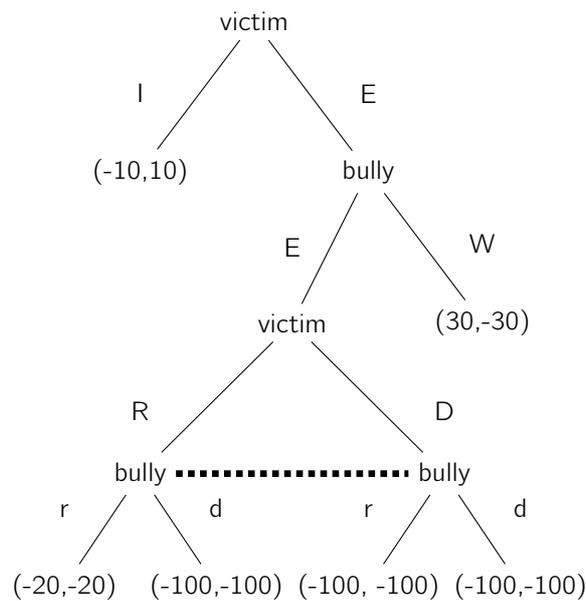
\begin{figure}[H]
	\centering
\begin{tikzpicture}
	% Stage 0
	\node{victim}
	child[level distance=2cm, sibling distance=3cm]{
		node[label={[yshift=0.5cm]above:{I}}]{(-10,10)}
	}
	child[level distance=2cm, sibling distance=3cm]{
		node[label={[yshift=0.5cm]above:{E}}]{bully}
		child[level distance=2cm, sibling distance=2cm]{
			node[label={[yshift=0.5cm]above:{E}}]{victim}
			child[level distance=2cm, sibling distance=4cm]{
				node[name=bully1,label={[yshift=0.3cm]above:{R}}]{bully}
				child[level distance=1.5cm, sibling distance=2cm]{
					node[label={[yshift=0.5cm]above:{r}}]{(-20,-20)}
				}
				child[level distance=1.5cm, sibling distance=2cm]{
					node[label={[yshift=0.5cm]above:{d}}]{(-100,-100)}
				}
			}
			child[level distance=2cm, sibling distance=4cm]{
				node[name=bully2,label={[yshift=0.3cm]above:{D}}]{bully}
				child[level distance=1.5cm, sibling distance=2cm]{
					node[label={[yshift=0.5cm]above:{r}}]{(-100, -100)}
				}
				child[level distance=1.5cm, sibling distance=2cm]{
					node[label={[yshift=0.5cm]above:{d}}]{(-100,-100)}
				}
			}
		}
		child[level distance=2cm, sibling distance=3cm]{
			node[label={[yshift=0.5cm]above:{W}}]{(30,-30)}
		}
	};
	
	% Dotted line between bully1 and bully2
	\draw[dotted, line width=2.5pt] (bully1)--(bully2);
\end{tikzpicture}
\caption{Sequential game structure of the model without control parameter. Stage I the victim chooses between strategy $E$ and $I$. Stage II the bully chooses between strategies E and W. Stage III is a simultaneous form game with incomplete information. The order of the strategies at stage III could be reversed without altering the outcome of the game.}
\label{tab:sgs}
\end{figure}

At stage III of the war of attrition game, potential outcomes include mutual retreat $(R,r)$ from the conflict, resulting in a negative emotional payoff $u^{r}_{i}(x,f)<0$ for both parties. Retreating at this stage yields a less damaging payoff for the bully compared to losing at stage two, due to certain third-party conditions regarding the disclosure of the outcome to the public. Conversely, the alternative outcomes involve strategies of mutual destruction $(D,d)$, $(D,r)$, and $(R,d)$, where both the bully and the victim endure severe negative consequences, with payoffs of $u^{d}_{i}(x,f)<u^{r}_{i}(x,f)<0$. The sequential conflict situation is outlined in Figure \ref{tab:sgs}, and the utility functions for each stage of the game are represented in Figure \ref{fig: main stable ut}.

\vspace{1cm}

In the following section, we present a detailed analysis of the model outlined above and provide strategic recommendations tailored to the victim's circumstances. We consider two scenarios: the Narcissist Bully Game without a control parameter and the Narcissist Bully Game with a control parameter.

\section{Analysis of the Narcissist Bully Game: Without Control Parameter}

\begin{table}[h!]
	\centering
\begin{tabular}{ccccc}
	\toprule
	\multirow{2}{*}{\rotatebox[origin=c]{90}{\textbf{ }}} & \textbf{Wr} & \textbf{Wd} & \textbf{Er} & \textbf{Ed}  \\
	\cmidrule{1-5}
	\textbf{IR}& -10,\underline{10} & -10,\underline{10} & \underline{-10} ,\underline{10}  & \underline{-10} \underline{10}  \\
	\textbf{ID} & -10,\underline{10} & -10,\underline{10}  & \underline{-10} ,\underline{10}  & \underline{-10} ,\underline{10}  \\
	\textbf{ER} & \underline{30},-30 & \underline{30},-30 & -20,\underline{-20} & -100,-100  \\
	\textbf{ED} & \underline{30},\underline{-30}  & \underline{30},\underline{-30} & -100,-100 & -100,-100  \\
	\bottomrule
\end{tabular}
\caption{Pure strategy Nash equilibria of the narcissist bully game without control parameter.}
\label{tab:psNEnoF}
\end{table}

\

From the narcissist bully game without control parameter in Figure \ref{tab:sgs} we obtain Table \ref{tab:psNEnoF} representing all pure Nash equilibria. The set of pure Nash equilibria is a set of strategies denoted by $$E^{NE}=\{(IR, Er), (Er, Ed), (ED,Wr),(ED,Wd)\}.$$

\

\begin{table}[h!]
	\centering
	\begin{tabular}{ccccc}
		\toprule
		\multirow{2}{*}{\rotatebox[origin=c]{90}{\textbf{ }}} & \textbf{Er} & \textbf{Ed}  \\
		\cmidrule{0-2}
		\textbf{ER} & \underline{-20},\underline{-20} & \underline{-100},-100  \\
		\textbf{ED} & -100,\underline{-100}& \underline{-100},\underline{-100} \\
		\bottomrule
	\end{tabular}
	\caption{Subgame perfect Nash equilibria of the war of attrition stage of the narcissist bully game without control parameter.}
	\label{tab:spNE 1}
\end{table}

From stage III in Figure \ref{tab:sgs} we further obtain the Table \ref{tab:spNE 1} representing all subgame perfect Nash equilibria of the war of attrition game. The subgame perfect Nash equilibria is a set of strategies denoted by $$E^{SPNE}=\{(ER, Er), (ED, Ed)\}.$$

\

\begin{figure}[h!]
	\centering
	\begin{subfigure}{0.45\textwidth}
		\centering
		\begin{tikzpicture}
			% Stage 0
			\node{victim}
			child[level distance=2cm, sibling distance=3cm]{
				node[label={[yshift=0.5cm]above:{I}}]{(-10,10)}
			}
			child[level distance=2cm, sibling distance=3cm,red]{
				node[label={[yshift=0.5cm]above:{E}}]{\textcolor{black}{bully}}
				child[level distance=2cm, sibling distance=2cm,red]{
					node[label={[yshift=0.5cm]above:{E}},red]{(-20,-20)}
				}
				child[level distance=2cm, sibling distance=3cm,black]{
					node[label={[yshift=0.5cm]above:{W}}]{(30,-30)}
				}
			};
		\end{tikzpicture}
		\caption{Sequential Game Stage II: SPNE $(ER, Er)$}
		\label{tab:spNE_ER,Er}
	\end{subfigure}%
	\hfill
	\begin{subfigure}{0.45\textwidth}
		\centering
		\begin{tikzpicture}
			% Stage 0
			\node{victim}
			child[level distance=2cm, sibling distance=3cm]{
				node[label={[yshift=0.5cm]above:{I}}]{(-10,10)}
			}
			child[level distance=2cm, sibling distance=3cm,red]{
				node[label={[yshift=0.5cm]above:{E}}]{\textcolor{black}{bully}}
				child[level distance=2cm, sibling distance=2cm,black]{
					node[label={[yshift=0.5cm]above:{E}}]{(-100, -100)}
				}
				child[level distance=2cm, sibling distance=3cm,red]{
					node[label={[yshift=0.5cm]above:{W}}]{(30,-30)}
				}
			};
		\end{tikzpicture}
		\caption{Sequential Game Stage II: SPNE $(ED, Ed)$}
		\label{tab:spNE_ED,Ed}
	\end{subfigure}
\caption{Analysis of subgames}
		\label{tab:spNE_ED,Ed both games}
\end{figure}
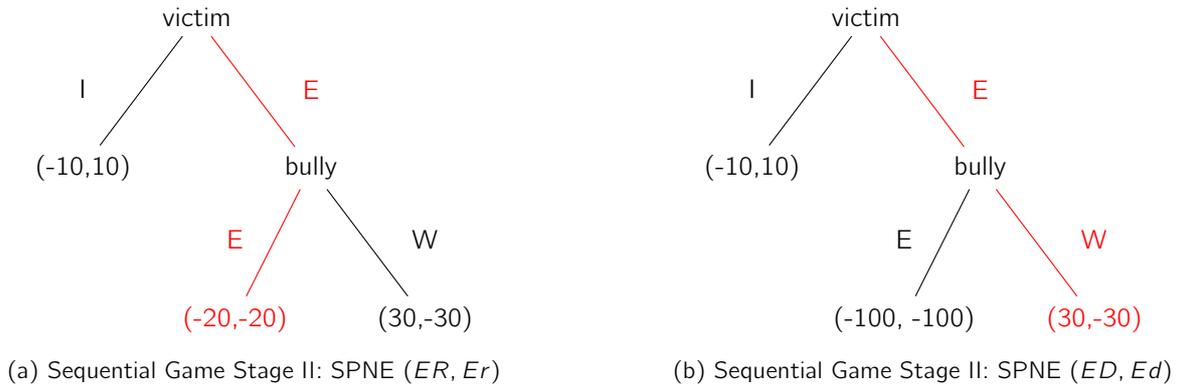
Consider the war of attrition stage of the game. We now analyze the subgame perfect Nash equilibria in the subgames using backward induction as shown in Figure \ref{tab:spNE_ED,Ed both games}. The subgames of both equilibrium strategies $(ER,Er)$ and $(ED,Ed)$ are evaluated. 

First we analyze $(ER,Er)$ with outcome $(-20,-20)$ as shown in Table \ref{tab:spNE 1}. We analyze it in the sequential structure of the model presented in Figure \ref{tab:spNE_ER,Er} using backward induction. Since at stage II of the sequential game, the bully prefers strategy $E$ over $W$ it follows that the victim will choose $E$ at stage I. Hence $(ER,Er)$  is an equilibrium strategy. 

Next, consider the strategy $(ED,Ed)$ with outcome $(-100,-100)$ presented in Table \ref{tab:spNE 1}.  We analyze it in the sequential structure of the model as shown in Figure \ref{tab:spNE_ER,Er} using backward induction. Since at stage II of the sequential game, the bully prefers strategy $W$ over $E$ it follows that the victim will choose $E$ at stage I. Hence $(ED,Ed)$ is an equilibrium strategy. The preceding analysis of $E^{SPNE}=\{(ER, Er), (ED, Ed)\}$ in Figure \ref{tab:spNE_ED,Ed} suggests that both SPNE equilibrium strategies are also equilibrium strategies of the sequential game. 

This analysis reveals that the victim’s willingness to escalate the situation, coupled with the credible threat of mutual detriment at the war of attrition stage, supports the existence of a strong deterrent. Given the robustness of this credible threat, we anticipate an additional equilibrium strategy $(E,W)$ with the outcome $(30,-30)$. To resolve any remaining ambiguity regarding equilibrium selection, we next introduce a control variable to refine and clarify the final equilibrium selection.

\section{Analysis of the Narcissist Bully Game: With Control Parameter}

\begin{figure}[h!]
	\centering
	\includegraphics[scale = 0.36]{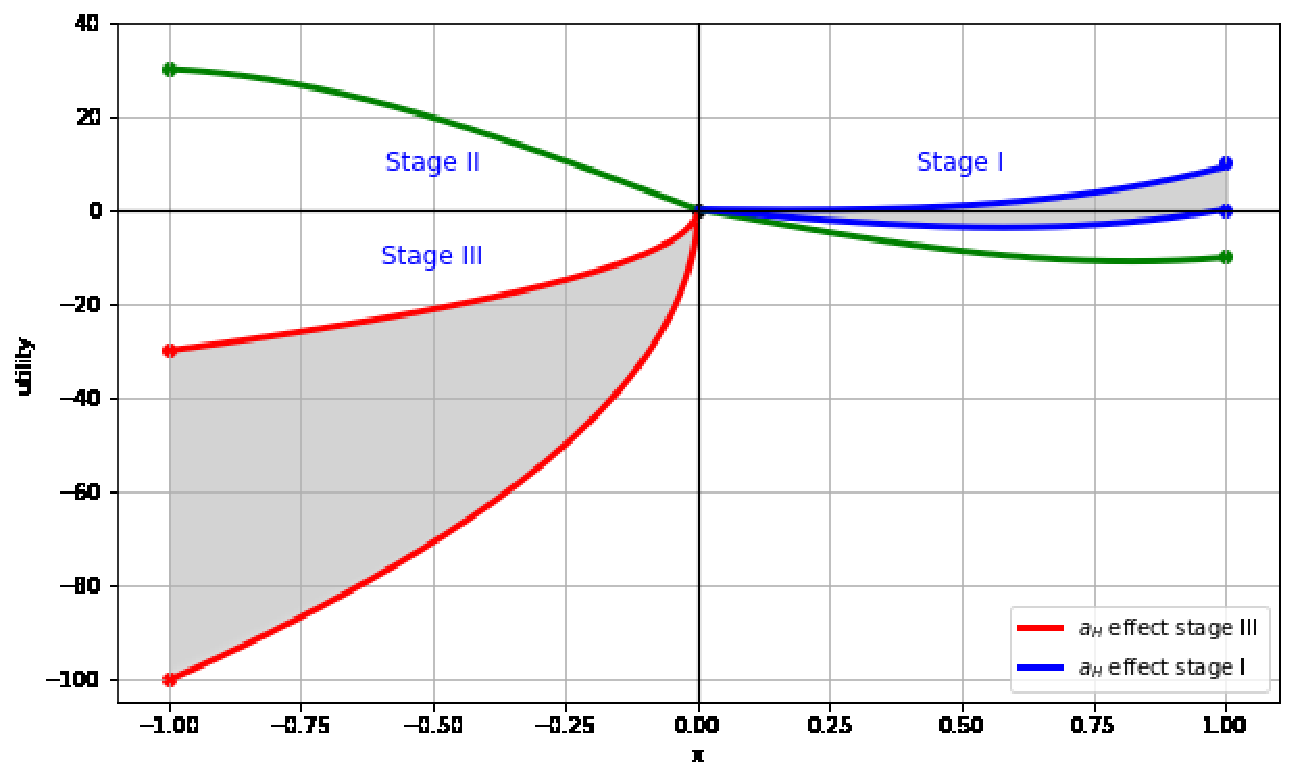}
	\includegraphics[scale = 0.36]{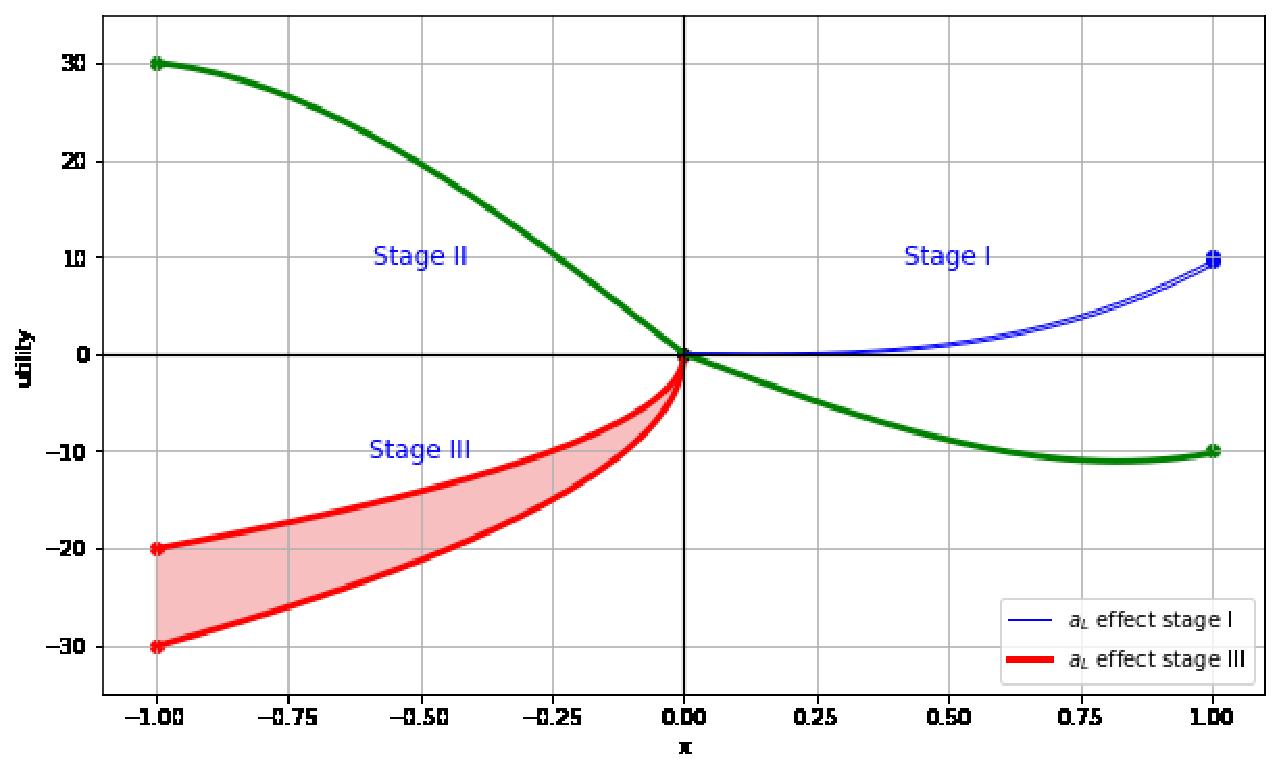}
	\caption{Narcissist Bully Game with low and high level Control Parameter: On the right, the graph shows the effect of low levels of control affects the bully's utility. On the left, the graph shows how high levels of control affect the utility payoffs of the bully.}
	\label{tab:main model control}
\end{figure}

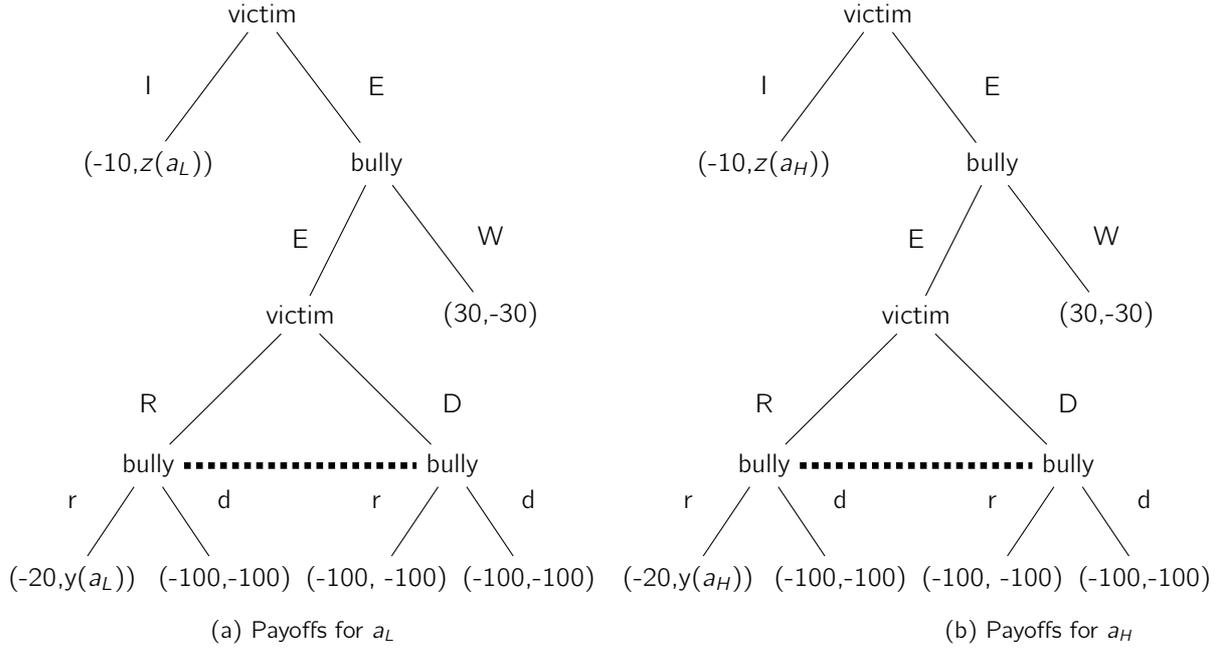
\begin{figure}[H]
	\defaultfigsize
	\hspace{0\textwidth} % Adjust the hspace value as needed
	\begin{subfigure}{0.5\textwidth}
		\begin{tikzpicture}[scale=1.0]
			% Stage 0
			\node{victim}
			child[level distance=2cm, sibling distance=3cm]{
				node[label={[yshift=0.5cm]above:{I}}]{(-10,$z(a_{L})$)}
			}
			child[level distance=2cm, sibling distance=3cm]{
				node[label={[yshift=0.5cm]above:{E}}]{bully}
				child[level distance=2cm, sibling distance=2cm]{
					node[label={[yshift=0.5cm]above:{E}}]{victim}
					child[level distance=2cm, sibling distance=4cm]{
						node[name=bully1,label={[yshift=0.3cm]above:{R}}]{bully}
						child[level distance=1.5cm, sibling distance=2cm]{
							node[label={[yshift=0.5cm]above:{r}}]{(-20,y($a_{L}$))}
						}
						child[level distance=1.5cm, sibling distance=2cm]{
							node[label={[yshift=0.5cm]above:{d}}]{(-100,-100)}
						}
					}
					child[level distance=2cm, sibling distance=4cm]{
						node[name=bully2,label={[yshift=0.3cm]above:{D}}]{bully}
						child[level distance=1.5cm, sibling distance=2cm]{
							node[label={[yshift=0.5cm]above:{r}}]{(-100, -100)}
						}
						child[level distance=1.5cm, sibling distance=2cm]{
							node[label={[yshift=0.5cm]above:{d}}]{(-100,-100)}
						}
					}
				}
				child[level distance=2cm, sibling distance=3cm]{
					node[label={[yshift=0.5cm]above:{W}}]{(30,-30)}
				}
			};
			% Dotted line between bully1 and bully2
			\draw[dotted, line width=2.5pt] (bully1)--(bully2);
		\end{tikzpicture}
			\caption{Payoffs  for $a_{L}$}
	\label{fig:twofigures sub 1}
	\end{subfigure}%
	\hspace{0.0\textwidth} % Adjust the hspace value as needed
	\begin{subfigure}{0.7\textwidth}
		\begin{tikzpicture}[scale=1.0]
			% Stage 0
			\node{victim}
			child[level distance=2cm, sibling distance=3cm]{
				node[label={[yshift=0.5cm]above:{I}}]{(-10,$z(a_{H})$)}
			}
			child[level distance=2cm, sibling distance=3cm]{
				node[label={[yshift=0.5cm]above:{E}}]{bully}
				child[level distance=2cm, sibling distance=2cm]{
					node[label={[yshift=0.5cm]above:{E}}]{victim}
					child[level distance=2cm, sibling distance=4cm]{
						node[name=bully1,label={[yshift=0.3cm]above:{R}}]{bully}
						child[level distance=1.5cm, sibling distance=2cm]{
							node[label={[yshift=0.5cm]above:{r}}]{(-20,y($a_{H}$))}
						}
						child[level distance=1.5cm, sibling distance=2cm]{
							node[label={[yshift=0.5cm]above:{d}}]{(-100,-100)}
						}
					}
					child[level distance=2cm, sibling distance=4cm]{
						node[name=bully2,label={[yshift=0.3cm]above:{D}}]{bully}
						child[level distance=1.5cm, sibling distance=2cm]{
							node[label={[yshift=0.5cm]above:{r}}]{(-100, -100)}
						}
						child[level distance=1.5cm, sibling distance=2cm]{
							node[label={[yshift=0.5cm]above:{d}}]{(-100,-100)}
						}
					}
				}
				child[level distance=2cm, sibling distance=3cm]{
					node[label={[yshift=0.5cm]above:{W}}]{(30,-30)}
				}
			};
			
			% Dotted line between bully1 and bully2
			\draw[dotted, line width=2.5pt] (bully1)--(bully2);
		\end{tikzpicture}
		\caption{Payoffs  for $a_{H}$}
	\label{fig:twofigures sub 2}
	\end{subfigure}
	\caption{Payoffs with low $a_{L}$ and high $a_{H}$ levels of fear.}
	\label{fig:twofigures}
\end{figure}

We consider the effect of introducing a control variable into the narcissistic bully game. Two levels of control are considered: (i) Low control level ($a_{L}$) ranging from $0.4 \leq a_{L} \leq 0.9$. In fact, when $a_{L} < 0.4$, inducing control has a negligible effect on the bully's preferences for successful or unsuccessful bullying. The model is then similar to the one shown in Figure \ref{tab:sgs}. (ii) We also consider a high level of induced control $(a_{H})$ ranging from $0.9 < a_{H} \leq 10$. The control-inducing models are represented in Figure \ref{fig:twofigures}. In the low level control model (Figure \ref{fig:twofigures sub 1}), the bully's payoff in stage I is denoted by $z(a_{L})$, and his stage III payoff is denoted by $y(a_{L})$. In the high level of control model (Figure \ref{fig:twofigures sub 2}), the bully's payoff in stage I is denoted by $z(a_{H})$, and his stage III payoff is denoted by $y(a_{H})$. The utility functions and payoffs are shown in Figure \ref{tab:main model control}.
\

\

For low levels of fear $0.4\leq a_{L}\leq 0.9$ we have stage I payoffs $z(a_{L}):=\{ \underline{u_{b}}(x;0.9)<u_{b}(x;a_{L})< 10:  0\leq x\leq 1 \text{ and } 0.4\leq a_{L}\leq 0.9\}$.
The strict inequalities arise from evaluating the function \(u_{b}(x;a_{L}) = 2.25x^3 - a_{L}x\) for \(0 < x < 1\), at the extreme points of $0.4\leq a_{L}\leq 0.9$.  We need to determine the critical points and check the endpoints of the given interval. $u_{b}(x_{min} ;a_{L}=0.4) = 0$ where $x_{min}=0$. $u_{b}(x^{*} ;a_{L}=0.4)  \approx -0.033$, where critical point $x^{*}=\sqrt{\frac{8}{135}} \approx 0.279$. $u_{b}(x_{max} ;a_{L}=0.4) = 1.85$ where $x_{max}=1$. $u_{b}(x_{min} ;a_{L}=0.9)  = 0$ where $x_{min}=0$. $u_{b}(x^{*} ;a_{L}=0.9)  \approx -0.086$, where critical point $x^{*}=\sqrt{\frac{2}{15}} \approx 0.327$. $u_{b}(x_{max} ;a_{L}=0.4) = 1.35$ where $x_{max}=1$. 
Similar calculations show that for low levels of fear $0.4\leq a_{L}\leq 0.9$ we have stage III payoffs $y(a_{L}):=\{-30\leq u_{b}(x;a_{L})\leq -20:  -1\leq x\leq 0 \text{ and } 0.4\leq a_{L}\leq 0.9\}$, where $u_{b}(x;a_{H})=-5.8  \sqrt(-30 a_{L}  x)$ for $-1\leq x\leq 0.$

\begin{table}[h!]
	\centering
	\begin{tabular}{ccccc}
		\toprule
		\multirow{2}{*}{\rotatebox[origin=c]{90}{\textbf{ }}} & \textbf{Wr} & \textbf{Wd} & \textbf{Er} & \textbf{Ed}  \\
		\cmidrule{1-5}
		\textbf{IR}& $-10,\underline{z(a_{L})}$ & $-10,\underline{z(a_{L})}$ & $-10,\underline{z(a_{L})} $& $-10,\underline{z(a_{L})}$ \\
		\textbf{ID}& $-10,\underline{z(a_{L})}$ & $-10,\underline{z(a_{L})}$ & $-10,\underline{z(a_{L})} $& $-10,\underline{z(a_{L})}$ \\
		\textbf{ER} & \underline{30},-30 & \underline{30},-30 & $-20, \underline{y(a_{L})}$ & -100,-100  \\
		\textbf{ED} & \underline{30},\underline{-30}  & \underline{30},\underline{-30} & -100,-100 & -100,-100  \\
		\bottomrule
	\end{tabular}
	\caption{Pure Strategy Nash Equilibria: Case $-20<y(a_{L})<-30$ and $10>z(a_{L})>0$.}
	\label{tab:psNE case aL}
\end{table}

We analyze the  $a_{L}$ case presented in Figure \ref{fig:twofigures sub 1}. Consider the bully's stage I payoff $10>z(a_{L})>0$ for successful bullying $x=1$ and stage III payoff  $-20<y(a_{L})<-30$ for unsuccessful bullying $x=-1$. The pure strategy Nash equilibria are shown in Table \ref{tab:psNE case aL}. 

From Table \ref{tab:psNE case aL} we obtain the pure Nash equilibria for $a_{L}$  is a set $$E^{NE}=\{(IR, Er), (Ir, Ed), (ID,Er),(ID,Ed), (ED,Wr), (ED,Wd)\}$$

\

\begin{table}[h!]
	\centering
	\begin{tabular}{ccccc}
		\toprule
		\multirow{2}{*}{\rotatebox[origin=c]{90}{\textbf{ }}} & \textbf{Er} & \textbf{Ed}  \\
		\cmidrule{0-2}
		\textbf{ER} & \underline{-20},$\underline{y(a_{L})}$& \underline{-100},-100  \\
		\textbf{ED} & -100,\underline{-100}& \underline{-100},\underline{-100} \\
		\bottomrule
	\end{tabular}
	\caption{SPNE: War of Attrition: $y(a_{L})$}
	\label{tab:attrition NE aL}
\end{table}

From Table \ref{tab:attrition NE aL} we obtain the Subgame Pure Nash equilibria $$E^{SPNE}=\{(ER, Er), (ED, Ed)\}.$$

\

For high levels of fear $0.9< a_{H}\leq 10$ we have stage I payoffs $z(a_{H}):=\{u_{b}(x;a_{H}=10)\leq u_{b}(x;a_{H})<u_{b}(x;a_{H}=0.9):  0\leq x\leq -1 \text{ and } 0.9< a_{L}\leq 10\}$. By calculations, we observe that $u_{b}(x;a_{H}=10)<0$ every where except at the endpoints of $0.9< a_{H}\leq 10$  where $u_{b}(x;a_{H}=0.9)=u_{b}(x;a_{H}=10)=0.$ A critical point arises at $x^{*}=0.577$ at which $u_{b}(x;a_{H}=10)=-3.85$. For high levels of fear $0.9< a_{H}\leq 10$ we have stage III payoffs $y(a_{H}):=\{-100\leq u_{b}(x;a_{H})<-30:  -1\leq x\leq 0 \text{ and } 0.9< a_{L}\leq 10\}$, where $u_{b}(x;a_{H})=5.8\sqrt{30(-a_{H}x}$ for $-1\leq x\leq 0$.

\

We analyze the  $a_{H}$ case presented in Figure \ref{fig:twofigures sub 2}. Consider the bully's stage I payoff $z(a_{L})=0$ for successful bullying $x=1$ and stage III payoff  $-30<y(a_{L})<-100$ for unsuccessful bullying $x=-1$. The pure strategy Nash equilibria are shown in Table \ref{tab:psNE cas aH}. 

\

\begin{table}[h!]
	\centering
	\begin{tabular}{ccccc}
		\toprule
		\multirow{2}{*}{\rotatebox[origin=c]{90}{\textbf{ }}} & \textbf{Wr} & \textbf{Wd} & \textbf{Er} & \textbf{Ed}  \\
		\cmidrule{1-5}
		\textbf{IR}& $-10,\underline{z(a_{H})}$ & $-10,\underline{z(a_{H})}$ & $\underline{-10},\underline{z(a_{H})} $& $\underline{-10},\underline{z(a_{H})}$ \\
		\textbf{ID}& $-10,\underline{z(a_{H})}$ & $-10,\underline{z(a_{H})}$ & $\underline{-10},\underline{z(a_{H})} $& $\underline{-10}\underline{z(a_{H})}$ \\
		\textbf{ER} & \underline{30},\underline{-30}& \underline{30},\underline{-30}& $-20, y(a_{H})$ & -100,-100  \\
		\textbf{ED} & \underline{30},\underline{-30}  & \underline{30},\underline{-30} & -100,-100 & -100,-100  \\
		\bottomrule
	\end{tabular}
	\caption{Pure Strategy Nash Equilibria: Case $-20<y(a_{H})<-30$.}
	\label{tab:psNE cas aH}
\end{table}

\

From Table \ref{tab:attrition NE aH} we obtain the pure Nash equilibria for $a_{L}$  is a set $$E^{NE}=\{(IR, Er), (Ir, Ed), (ID,Er),(ID,Ed), (ER,Wr), (ER,Wd), (ED,Wr), (ED,Wd)\}.$$

\begin{table}[h!]
	\centering
	\begin{tabular}{ccccc}
		\toprule
		\multirow{2}{*}{\rotatebox[origin=c]{90}{\textbf{ }}} & \textbf{Er} & \textbf{Ed}  \\
		\cmidrule{0-2}
		\textbf{ER} & \underline{-20},$\underline{y(a_{H})}$& \underline{-100},-100  \\
		\textbf{ED} & -100,\underline{-100}& \underline{-100},\underline{-100} \\
		\bottomrule
	\end{tabular}
	\caption{SPNE: War of Attrition: $y(a_{H})$}
	\label{tab:attrition NE aH}
\end{table}

From Table \ref{tab:attrition NE aH} we obtain the Subgame Pure Nash equilibria $$E^{SPNE}=\{(ER, Er), (ED, Ed)\}.$$

\

\begin{figure}[h!]
	\centering
	\begin{subfigure}{0.45\textwidth}
		\centering
		\begin{tikzpicture}
			% Stage 0
			\node{victim}
			child[level distance=2cm, sibling distance=3cm]{
				node[label={[yshift=0.5cm]above:{I}}]{(-10,$z(a_{H})$)}
			}
			child[level distance=2cm, sibling distance=3cm, red]{
				node[label={[yshift=0.5cm]above:{E}}]{\textcolor{black}{bully}}
				child[level distance=2cm, sibling distance=2cm, black]{
					node[label={[yshift=0.5cm]above:{E}}]{(-20,$y(a_{H})$)}
				}
				child[level distance=2cm, sibling distance=3cm, red]{
					node[label={[yshift=0.5cm]above:{W}}]{(30,-30)}
				}
			};
		\end{tikzpicture}
		\caption{SPNE: Sequential Game: $a_{L}$, $(ER,Er)$}
		\label{tab:spNE_Rr}
	\end{subfigure}%
	\hfill
	\begin{subfigure}{0.45\textwidth}
		\centering
		\begin{tikzpicture}
			% Stage 0
			\node{victim}
			child[level distance=2cm, sibling distance=3cm]{
				node[label={[yshift=0.5cm]above:{I}}]{(-10,10)}
			}
			child[level distance=2cm, sibling distance=3cm, red]{
				node[label={[yshift=0.5cm]above:{E}}]{\textcolor{black}{bully}}
				child[level distance=2cm, sibling distance=2cm, black]{
					node[label={[yshift=0.5cm]above:{E}}]{(-100, -100)}
				}
				child[level distance=2cm, sibling distance=3cm, red]{
					node[label={[yshift=0.5cm]above:{W}}]{(30,-30)}
				}
			};
		\end{tikzpicture}
		\caption{SPNE: Sequential Game: $a_{L}$, $(ED,Ed)$}
		\label{tab:spNE_Dr}
	\end{subfigure}
	\caption{SPNE: Sequential Game: $a_{L}$ and $a_{H}$.}
	\label{fig:twofigures_aL_and_aH}
\end{figure}
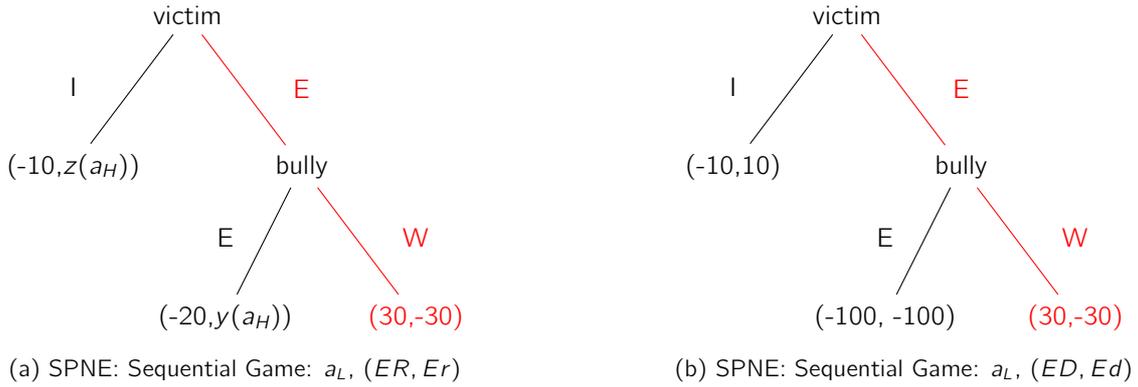

We now analyze the sequential game equilibria for both $E^{SPNE}=\{(ER, Er), (ED, Ed)\}$ in both cases $a_{L}$ and $a_{H}$. The analysis for $a_{L}$ results in the same equilibrium strategies as in the original scenario suggesting that low levels of fear $a_{L}$ have no effect on equilibrium selection as can be seen form the analysis represented in Figure \ref{tab:spNE_ED,Ed both games}.  Figure \ref{fig:twofigures_aL_and_aH} represents the case of the high level of induced fear, $a_{H}$ which we now analyze. Since $-30<y(a_{H})<-100$ in $(ER, Er)$ the bully chooses strategy $W$ over $E$ in stage III. Knowing that, the victim chooses $E$ over $I$ in stage I. Similarly, for $(ED, Ed)$ the bully chooses strategy $W$ over $E$ in stage III. Knowing this, the victim chooses $E$ over $I$. The analysis suggests that a high level of fear reduces the number of equilibria to a unique sequential game equilibrium, which is to the advantage of the victim. 

It follows from the analysis of subgame perfect Nash equilibria \ref{fig:twofigures_aL_and_aH} that $$E^{SPNE}=\{(ER, Wr), (ER, Wd)\}\iff \{(E,W)\},$$ hence a reduction of equilibrium selection to the unique equilibrium strategy $(E,W)$ with associated payoffs $(30,-30)$.

\section{Conclusion}

Instilling doubt in narcissistic bosses can be an effective strategy to curb their bullying behavior. While narcissistic individuals often project confidence and self-assurance, certain psychological triggers—particularly those that challenge their ego, control, or reputation—can make them feel vulnerable and unsettled. For example, narcissists are highly protective of their public image and fear reputational damage, making transparent performance evaluations and public feedback powerful methods to counteract their dominance \cite{Kernis2005}. They also tend to feel anxious in environments where their control is diminished, such as in shared decision-making or team-based initiatives, which limit their influence and challenge their authority \cite{Campbell2009}. Furthermore, narcissists are sensitive to social rejection and isolation, as they rely on validation and admiration from others; fostering team cohesion can weaken the narcissist's influence within the group and reduce their control over individuals \cite{Leary2000}. Unpredictable responses from colleagues—such as calm assertiveness or unexpected confrontation—can destabilize narcissists, who often expect submissive reactions and may feel threatened when their intimidation tactics are ineffective \cite{Morf2001}. Finally, regular accountability measures, such as peer reviews and structured performance assessments, introduce transparency that curtails narcissists’ ability to manipulate outcomes, promoting an environment that prioritizes merit over their self-serving tendencies \cite{Brunell2017}.

Our analysis underscores the power of assertive, strategically calculated responses for individuals targeted by workplace bullying, particularly in the context of narcissistic aggression. In situations free from the influence of fear, the most effective approach for a victim is escalation (strategy $E$), employing the credible threat of mutual detriment—a threat whose cost outweighs that of tolerating further bullying. The presence of two sequentially perfect Nash equilibria within this framework highlights the pivotal role of the victim's credibility in sustaining this threat across both potential outcomes. However, when fear factors into the model, the victim can simplify the dynamics to a single, advantageous equilibrium by instilling doubt and apprehension in the narcissist’s mind. Under conditions of heightened fear, strategy $E$ emerges as the singular optimal path, thereby shifting the balance decisively in the victim’s favor.

These findings underscore broader applicability beyond individual response strategies, offering insights for institutions seeking to improve workplace conditions and address narcissistic bullying more effectively. In both academic and professional settings, victims of narcissistic bullying benefit significantly from structural mechanisms that bolster their credibility and resilience. For example, transparent performance evaluations, 360-degree feedback, and formal anti-bullying policies introduce layers of accountability, empowering faculty and staff to address issues confidently and without fear of retaliation, thereby curbing unchecked bullying behaviors \cite{Keashly2010}. Additionally, initiatives such as structured mentorship programs, rotation of leadership roles, and anonymous peer reviews foster a fair and balanced work environment, reducing opportunities for narcissistic individuals to exploit hierarchical dynamics \cite{Johnson2015, Bornmann2008}. Team-based recognition and reward systems further shift the focus from individual dominance to collective achievement, creating an inclusive culture that discourages bullying and enhances morale \cite{Grijalva2014}.

By implementing these strategies, institutions can create a culture that not only mitigates the effects of bullying but also proactively deters narcissistic behaviors. This dual approach—combining assertive responses from individuals with structural improvements—supports a workplace culture rooted in respect, transparency, and collaboration, ultimately building a healthier, more resilient academic and professional community.

\end{document}